\documentclass[journal]{IEEEtran}
\usepackage{multirow}
\usepackage{subfigure}
\usepackage{graphicx}
\usepackage{amsmath}
\usepackage{amsfonts}
\usepackage{algorithm}
\usepackage{algorithmic}
\usepackage{makecell}
\usepackage{booktabs}
\setcellgapes{3pt}

\setcellgapes{3pt}

\usepackage[justification=centering]{caption}
\usepackage{color}

\usepackage{easyReview}

\usepackage{cite}
\usepackage{verbatim}

\ifCLASSINFOpdf

\else

\fi

\begin{document}

\title{Agentic AI Empowered  Multi-UAV Trajectory Optimization in Low-Altitude Economy Networks}

\author{Feibo Jiang, \textit{Senior Member, IEEE}, Li Dong, Xitao Pan, Kezhi Wang, \textit{Senior Member, IEEE}, and Cunhua Pan, \textit{Senior Member, IEEE}.
\thanks{This work was supported in part by the National Natural Science Foundation of China under Grant nos. 41904127 and 41874148.}

\thanks{Feibo Jiang (jiangfb@hunnu.edu.cn) is with Hunan Provincial Key Laboratory of Intelligent Computing and Language Information Processing, Hunan Normal University, Changsha, China.}
\thanks{Li Dong (Dlj2017@hunnu.edu.cn) is with Key Laboratory of Hunan Province for New Retail Virtual Reality Technology, Hunan University of Technology and Business, Changsha, China.}
\thanks{Xitao Pan (1837349760@hunnu.edu.cn) is with Hunan Provincial Key Laboratory of Intelligent Computing and Language Information Processing, Hunan Normal University, Changsha, China.}%

\thanks{Kezhi Wang (Kezhi.Wang@brunel.ac.uk) is with the Department of Computer Science, Brunel University London, UK.}
\thanks{Cunhua Pan (cpan@seu.edu.cn) is currently a full professor in Southeast University, China.}
}

%
%

\markboth{Submitted for Review}%
{Shell \MakeLowercase{\textit{et al.}}: Bare Demo of IEEEtran.cls for IEEE Journals}
%



\maketitle
\begin{abstract}

This paper proposes a novel Agentic Retrieval-augmented generation with Mamba-Attention Integrated Transformer (ARMAIT) framework for multi-Unmanned Aerial Vehicle (UAV) trajectory optimization. The framework is built upon Large Language Models (LLMs), incorporating Retrieval-Augmented Generation (RAG) empowered by Agentic AI and integrated with a UAV-specific knowledge base. Through the Agentic RAG, the LLM autonomously interprets high-level task requirements and identifies the key components necessary for trajectory optimization, including model inputs and outputs, network architecture, reward functions, and task constraints.
To support efficient modeling across different system scales, we introduce the Mamba-Attention Integrated Transformer (MAIT), a hybrid neural architecture that combines the long-range dependency modeling capability of attention mechanisms with the efficient temporal dynamic representation of Mamba. Furthermore, a Trajectory-Group Relative Policy Optimization (T-GRPO) method is proposed to achieve unified policy gradient optimization in both discrete and continuous trajectory spaces for MAIT training.
Extensive experimental results validate the feasibility and effectiveness of the proposed ARMAIT framework.

\end{abstract}

%
\IEEEpeerreviewmaketitle

\section{Introduction}
\label{sec:introduction}

%
%
%
%

With the continuous convergence of Information and Communication Technologies (ICT) and Unmanned Aerial Vehicle (UAV) technologies, Low-Altitude Economy Networks (LAENets) are gradually emerging as a critical new infrastructure to support future societal activities \cite{wang2025toward}. These networks operate primarily in low-altitude airspace below 3,000 meters, utilizing a variety of Low-Altitude Vehicles (LAVs), such as taxis, aerial base stations, and airborne charging air stations. These vehicles are capable of efficiently performing diverse tasks, including parcel transportation, daily goods delivery, communication coverage, and aerial monitoring. To enable such functionalities, LAENets require the deep integration of key service capabilities, including communication, computation, storage, and sensing. However, compared with traditional terrestrial communication networks, LAENets exhibit significant differences in control mechanisms and resource allocation strategies. The highly dynamic and unpredictable nature of their operating environments leads to increased system complexity and heightened vulnerability to security threats \cite{zhao2025generative}. Therefore, there is an urgent need for robust and intelligent system designs tailored to the specific characteristics and challenges of LAENets.

In LAENets, UAVs serve as essential airborne nodes and are widely deployed due to their high mobility and flexible deployment capabilities. UAVs can function as aerial base stations or relay nodes to extend the spatial coverage of cellular networks, thereby enabling communication access in remote or underserved areas. UAVs have also demonstrated significant potential in a variety of mission scenarios, such as medical supply delivery, daily logistics, and last-mile urban distribution \cite{betti2024uav}. The effectiveness of UAV task execution and the efficiency of their trajectory planning have a direct impact on the overall performance of the network. Therefore, developing efficient UAV scheduling and trajectory optimization mechanisms is a fundamental requirement for ensuring the stable and reliable operation of LAENets.

In recent years, Deep Learning (DL) and Reinforcement Learning (RL) methods have been widely applied to UAV trajectory planning tasks due to their powerful modeling capabilities for high-dimensional optimization problems. DL approaches have shown remarkable performance in feature extraction and policy learning \cite{liang2017text}. However, conventional fully connected neural network architectures typically assume fixed-length inputs, making them less suitable for scenarios with dynamically varying numbers of devices and task scales. To address this limitation, researchers have proposed extended models based on Recurrent Neural Networks (RNNs), such as Long Short-Term Memory (LSTM) networks and Pointer Networks (PNs), which are capable of handling sequential data and extracting heterogeneous features. Nonetheless, although RNNs can model temporal dependencies, they often suffer from gradient vanishing and memory bottlenecks when dealing with long sequences, which limits their ability to capture global dependencies among distant nodes \cite{salehinejad2017recent}.

Meanwhile, RL-based methods such as Deep Q-Networks (DQN) and Deep Deterministic Policy Gradient (DDPG) have also been extensively used for optimizing UAV deployment and trajectory planning. These approaches leverage state-action space modeling to achieve joint optimization of energy consumption and performance. However, RL methods often face challenges such as slow policy convergence and training instability in high-dimensional continuous action spaces. Moreover, their generalization ability tends to degrade when dealing with dynamic changes in the task complexity, limiting their applicability in large-scale and real-time trajectory planning scenarios \cite{lyle2022learning}.

To overcome the aforementioned limitations, the Transformer architecture has been increasingly adopted in UAV trajectory planning tasks due to its inherent advantages in handling variable-length inputs, modeling global dependencies, and enabling efficient parallel computation \cite{jiang2025path}. The self-attention mechanism in Transformer models allows the network to directly capture dependencies between arbitrary positions in the input sequence without relying on sequential order, thereby enhancing the global consistency of path generation. Moreover, the Transformer structure is highly compatible with varying input lengths and structures, making it well-suited for typical LAENets where the number of UAVs and devices is not fixed \cite{vaswani2017attention}.

However, the standard Transformer relies on a global self-attention mechanism, whose computational complexity grows quadratically with the input sequence length. This results in substantial resource consumption when handling large numbers of devices or multi-UAV cooperative tasks, making it difficult to meet the requirements of edge deployment and real-time trajectory planning. In addition, the Transformer lacks the ability to continuously model state evolution, which limits its effectiveness in capturing the underlying dynamic physical patterns in UAV trajectories, such as variations in energy consumption, resource constraints, and spatial transitions \cite{chen2023contiformer}. To address these limitations, the recently proposed Mamba architecture offers a promising direction for enhancing the Transformer’s capability in continuous dynamic modeling \cite{gu2023mamba}. Mamba employs linear State Space Models (SSMs) to represent sequential data in a continuous manner, enabling efficient modeling of long-range dependencies and low-latency inference. Compared with traditional attention-based mechanisms, Mamba is inherently suited for long-sequence inputs and eliminates the need for costly attention matrix computations, significantly reducing memory and computational overhead. This makes it particularly advantageous for trajectory planning tasks in LAENets.

To advance the autonomous decision-making and trajectory optimization capabilities of multi-UAV systems in LAENets, this paper explores the integration of the Agentic Artificial Intelligence (AI) paradigm. By combining Large Language Models (LLMs), trajectory planning models, and RL algorithms, we construct an Agentic Retrieval-augmented generation with Mamba-Attention Integrated Transformer (ARMAIT) framework that unifies task comprehension, knowledge retrieval, model formulation and training, and trajectory generation. 
The main contributions of this study are summarized as follows:
\begin{enumerate}
	\item This work introduces the Agentic AI paradigm into multi-UAV trajectory planning. We design an intelligent framework that integrates LLMs with a UAV-specific knowledge base. Leveraging an Agentic Retrieval-Augmented Generation (RAG) mechanism, the LLM is empowered to autonomously interpret user intent from natural language task instructions, retrieve relevant UAV knowledge, and automatically generate the required model components, including input representations (e.g., states space, output objectives (e.g., actions space), model architecture (e.g., the ratio of Attention to Mamba layers), reward function definitions and task constraints. This mechanism establishes a closed-loop process from language understanding to task modeling, significantly enhancing the adaptability and generality of multi-UAV trajectory planning tasks.
	
	\item To address the trajectory optimization of UAV swarms of varying sizes, we propose a Mamba-Attention Integrated Transformer (MAIT) model. This model combines the spatial global dependency modeling strength of multi-head attention with the efficient long-sequence modeling capability of the Mamba architecture. MAIT supports parallel interaction across multiple UAV states while capturing temporal dynamics essential for trajectory planning. The UAV system states are encoded into structured input sequences, and the model architecture adapts the attention-to-Mamba layer ratio automatically based on the UAV fleet size, favoring attention layers for smaller-scale systems and Mamba layers for larger-scale ones. This design ensures strong generalization and high decision-making efficiency.
	
	\item We propose Trajectory-Group Relative Policy Optimization (T-GRPO), a policy-gradient RL algorithm tailored for multi-UAV trajectory generation. Building upon the original GRPO method, T-GRPO extends its applicability to both discrete (e.g., stop point sequences) and continuous (e.g., flight angles and distances) action spaces. The algorithm defines state and action spaces for UAV swarms as well as a global reward function. It introduces a group-based relative advantage function and policy divergence constraints to optimize the policy. This enables the stable training of the MAIT model for generating globally coordinated multi-UAV flight trajectories across diverse task scenarios.
\end{enumerate}

This paper is organized as follows: Section \ref{sec:related} reviews some related work. 
Section \ref{sec:UFO} describes the principle of the ARMAIT framework in detail. 
Section \ref{sec:experiments1}-\ref{sec:experiments2} illustrates the experiment results. Finally, Section \ref{sec:conclusion} concludes the paper and discusses future works. 

\section{Related Work}
\label{sec:related}

\subsection{Recurrent Neural Network-based Methods}
In \cite{8758355}, the authors introduced a novel design method for controllers, based on adaptive multilayer neural dynamics, which allowed multirotor UAVs to successfully carry out time-varying trajectory tracking tasks, even amidst unknown disturbances.
In \cite{9880930},  the authors explored fault-tolerant tracking control for networked fixed-wing UAVs against faults and communication delays, and creatively used double recurrent perturbation Fuzzy Neural Networks as a solution.
A LSTM-based model was proposed in \cite{2014lstm-CO}, which extracted the features of input nodes in order to generate UAV trajectories, while \cite{2015pointer} trained the PN to learn the mapping relationship between nodes, trajectories, and other combinatorial problems. The PN model was used by \cite{2022pn-uav} in a different way, using it to improve energy efficiency as well as generate optimal UAV trajectories. 

\subsection{Reinforcement Learning-based Methods}
Q-learning was used to optimize the waypoints and trajectory of UAVs in \cite{2020uav-qlearning} while the authors used an RL-based algorithm for trajectory control in order to optimize the system energy cost in \cite{2021muav-drl}. DDPG was utilized in \cite{2018uav-ddpg} to control the UAV and maximize energy efficiency, resulting in maximizing energy efficiency and achieving fair coverage for all users.
In \cite{10045775}, the authors formulated a data-driven approach, utilizing static output feedback control, grounded in inverse RL principles.
In \cite{9345436}, the authors employed adaptive and robust techniques, incorporating the DDPG, to manage the tracking control of quadrotors. The focus was on position control of the robotic arm, aiming to minimize its impact on quadrotor dynamics while adhering to the desired trajectory.

\subsection{\textcolor{black}{Transformer-based Methods}}

The learning efficiency of the RNN-based models is low because these models preclude parallelization\cite{2017transformer}. Moreover, RNN-based models may suffer from the gradient vanishment\cite{2021E2ECO}.
Meanwhile, RL-based methods suffer from low search efficiency in large-scale multi-UAV scenarios and encounter difficulties in achieving convergence during training in dynamic environments\cite{2021-mec-distributed}. The RL with transformer models can effectively address these issues.

\textcolor{black}{Transformers and their variants have proven effective in solving complex optimization problems in UAV-assisted applications.
	In multi-UAV area coverage, Chen et al. \cite{10423879} employed a Transformer encoder within their T-MARL algorithm to adapt to variable numbers of UAVs and users, effectively processing heterogeneous network state information for scalable operations.
	For UAV trajectory planning to minimize Age of Information (AoI), Zhu et al. \cite{9892691} utilized a Transformer encoder-decoder architecture (TWA*) to translate the entire UAV-IoT system state into an optimized visiting order, achieving good generalization and computational efficiency for the problem.
	In Multi-Agent Reinforcement Learning (MARL) for UAV communication, Feng et al. \cite{10918783} embedded a lightweight Transformer into the critic network of each UAV BS in their EDT-MARL framework. 
	Furthermore, in UAV-to-UAV visual tracking, Wang et al.\cite{10846255} used a Vision Transformer (ViT)-based model for robust visual feature learning in the main tracker, complemented by a separate Transformer-based module for predicting target motion trajectories from sequential data, enhancing tracking in highly dynamic environments.}

However, existing UAV trajectory planning methods are typically designed for specific system models and lack the flexibility to accommodate diverse scenarios and task requirements. To address this limitation, we propose the ARMAIT framework, which integrates LLMs and an Agentic RAG mechanism to interpret varying task demands and convert them into concrete design specifications. In addition, we develop the MAIT model, adaptable to UAV systems of different scales, and the T-GRPO training method, which supports both discrete and continuous trajectory planning. Our goal is to achieve a highly generalizable and task-adaptive multi-UAV trajectory planning solution.

\section{ARMAIT Framework}
\label{sec:UFO}

\begin{figure*}[htbp]
	\centering
	\includegraphics[width=17cm]{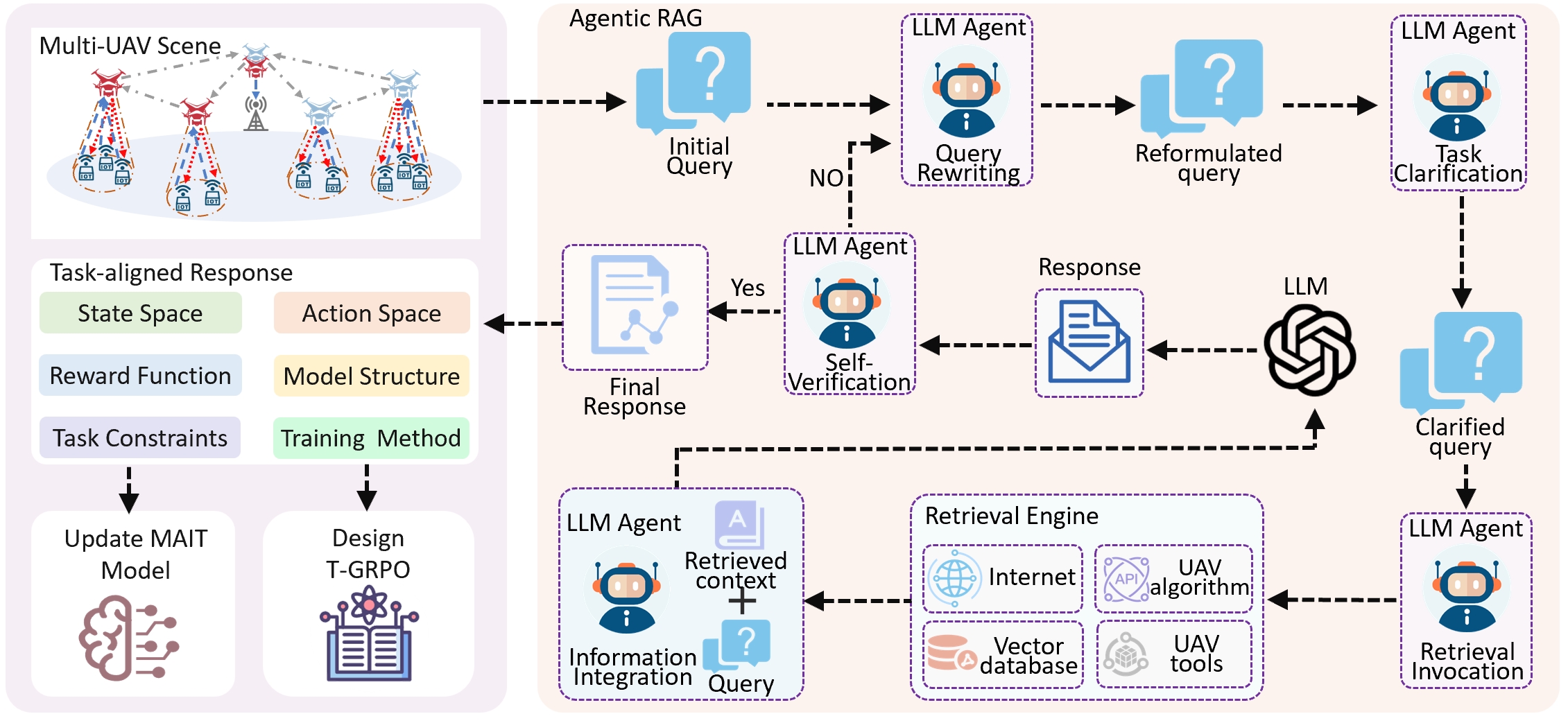}
	\caption{The workflow of the ARMAIT framework.}
	\label{fig:ARMAIT}
\end{figure*}

The ARMAIT framework comprises three core modules: the Agentic RAG task analysis module, the MAIT path generation model, and the T-GRPO optimizer. Specifically, the Agentic RAG module leverages an LLM integrated with a UAV domain knowledge base to automatically interpret task intent from natural language descriptions, infer modeling requirements, and generate model architectures and data interfaces for MAIT and T-GRPO. The MAIT model combines the global representation capabilities of the attention model with the efficient sequential modeling strengths of the Mamba model, enabling the efficient generation of coordinated trajectories for multiple UAVs. T-GRPO operates within an RL framework to jointly optimize policies over discrete and continuous action spaces, training the MAIT model to generate coordinated trajectories that satisfy multiple constraints. Together, these components form a closed-loop end-to-end intelligent system, enabling seamless transition from natural language understanding to multi-UAV trajectory planning. The workflow of the ARMAIT framework is shown in Fig. \ref{fig:ARMAIT}.

\subsection{Agentic RAG Task Analysis Module}

To effectively analyze diverse UAV path planning tasks, LLMs are increasingly employed due to their powerful natural language understanding and reasoning capabilities. However, current LLM-based task analysis often lacks sufficient domain-specific knowledge of UAV systems, leading to challenges such as inaccurate modeling and the generation of hallucinated content. These limitations hinder the ability of LLMs to generate precise, task-aligned planning strategies in real-world UAV applications.

RAG systems mitigate some of these challenges by incorporating a semantic retrieval module prior to generation. This mechanism expands user queries into context-rich prompts that are then processed by the LLM, effectively addressing issues related to hallucination and knowledge staleness. Nevertheless, conventional RAG systems still exhibit critical limitations when applied to UAV-related task analysis: (1) The user query cannot dynamically adapt to the evolving task context of UAVs, resulting in inconsistencies between the retrieved content and the generation objective; (2) The lack of procedural control prevents dynamic adjustment of the retrieval strategy or refinement of the query expression based on intermediate outputs, thereby inhibiting autonomous reflection and self-correction. These shortcomings substantially constrain the practical applicability of RAG systems in complex UAV task modeling and reasoning.

To overcome the limitations of static modeling and single-round inference in traditional RAG systems, this study introduces a novel Agentic RAG architecture for UAV task modeling and reasoning. This system integrates the core principles of Agentic AI, which endow LLMs with autonomous perception, reasoning, and proactive planning capabilities. By enabling dynamic decision-making and self-directed retrieval for complex UAV tasks, Agentic RAG establishes a closed-loop task-solving pipeline with autonomous reasoning chains for trajectory planning. The following outlines the operational workflow of Agentic RAG for analyzing UAV trajectory planning tasks:

\subsubsection{Initial Query Reception}
The system receives a natural language input from the user describing a UAV-related task (e.g., ``Please dispatch 10 UAVs to collect data from 100 IoT devices in a designated area, optimizing for energy consumption").

\subsubsection{Query Rewriting}
Upon understanding the task objectives and key variables, the LLM Agent semantically reconstructs the initial input. For example, the input may be reformulated as: ``Based on the criterion of minimizing energy consumption, construct a multi-UAV trajectory planning model using reinforcement learning to coordinate the flight paths of 10 UAVs for efficient data collection from 100 IoT devices in a target area."

\subsubsection{Task Clarification} 
The Agent determines whether additional background information is required, such as scenario models, flight altitude, or battery constraints. It then prompts the user to provide these parameters to complete the necessary prior knowledge. For example, the Agent may automatically ask: ``Please provide the geographical distribution of IoT devices, flight distance limitations, and the maximum payload capacity of each UAV for improved modeling accuracy."

\subsubsection{Contextual Retrieval}

The clarified query is submitted to a retrieval engine, which includes a vector database constructed from UAV trajectory planning literature, a tool library composed of various UAV-related models (such as UAV dynamics models and wireless channel models), an algorithm library containing diverse UAV-related algorithms (such as convex optimization algorithms and genetic algorithms), and API interfaces with direct access to the Internet.

\subsubsection{Information Integration and Generation}
The Agent integrates the retrieved information with the user’s original intent for downstream modeling. For instance, it may issue a prompt such as:
``Based on the retrieved knowledge, please generate a draft design of a multi-UAV trajectory optimization system. Include the definition of state space, action space, constraint condition, reward function, and recommend a suitable MAIT architecture and T-GRPO method."
This then triggers the LLM to perform generation and reasoning, resulting in an initial output, such as:
\begin{itemize}
	\item \emph{State space}: Position, battery level, remaining task load.

	\item \emph{Action space}: Discrete motion primitives.

	\item \emph{Reward function}: The total UAV energy consumption, including flight energy, hovering energy, and wireless charging energy.

	\item \emph{Model architecture}: A Transformer with 10 encoder layers, comprising 50\% Mamba layers and 50\% attention layers.

    \item \emph{Training method}: Discrete trajectory training with 8 independent rollouts.
    
    \item \emph{Task constraints}: The total energy consumption must not exceed the UAV’s maximum energy capacity, and the total amount of collected data must remain within its maximum storage capacity.
\end{itemize}
\subsubsection{Self-Verification and Evaluation}
The Agent semantically verifies the generated output by checking whether all essential parameters are included and whether the proposed design satisfies task constraints.
For example, the Agent may identify the absence of resource constraints. 

\subsubsection{Multi-Round Retrieval and Generation}
If the current output does not meet task objectives or lacks sufficient contextual grounding, the Agent restarts the query rewriting and retrieval cycle for another round of interaction. Otherwise, the finalized solution is returned to the MAIT model and the T-GRPO algorithm.

In summary, Agentic RAG overcomes the limitations of static and passive operations in traditional RAG systems by introducing an LLM Agent equipped with autonomous reasoning and feedback control capabilities. This paradigm establishes a generalizable framework for building intelligent UAV systems with task autonomy and lays a foundational infrastructure for applications such as multi-UAV scheduling and automated modeling. The system not only exhibits enhanced task comprehension and multi-source knowledge integration, but also enables complex task decomposition, self-optimization, and output verification. Consequently, it significantly improves the adaptability, accuracy, and robustness of LLMs in UAV scenarios involving complex modeling, decision support, and interactive intelligent generation. The workflow of the Agentic RAG is presented in \textbf{Algorithm \ref{alg:overview}}.

\begin{algorithm}
	\caption{Agentic RAG}
	\label{alg:overview}
\begin{algorithmic}[1]
	\REQUIRE Natural language task description.
	\ENSURE Task-aligned model architecture and training strategy.
	
	\STATE Receive the initial user query describing the UAV trajectory planning task.
	\STATE The LLM Agent semantically rewrites and enriches the query with task objectives and variables.
	\IF{critical parameters are missing}
	\STATE Prompt the user to supplement essential parameters.
	\ENDIF
	\STATE Submit the reformulated query to the retrieval engine.
	\STATE Integrate the retrieved knowledge with the user’s query.
	\STATE Generate an initial solution. 
	
	\STATE Perform self-verification of the generated solution.
	\IF{the solution is incomplete or suboptimal}
	\STATE Restart the query rewriting and retrieval cycle.
	\ELSE
	\RETURN the finalized solution to \textbf{Algorithm 2} and \textbf{Algorithm 3}.
	\ENDIF
\end{algorithmic}
\end{algorithm}

\subsection{Mamba-Attention Integrated Transformer}
\label{sec:encoder}

Integrating the Transformer architecture with trajectory optimization algorithms has emerged as a cutting-edge approach to enhancing the intelligent scheduling capabilities of UAV swarms. 
 A hybrid trajectory optimization framework is constructed by fusing the Mamba architecture with the self-attention mechanism of Transformer models. This framework preserves the global modeling capability of Transformers while introducing an efficient mechanism for dynamic modeling. Within this architecture, the attention module captures global dependencies and dynamic interactions among input nodes, whereas the Mamba module effectively models the continuity of device mobility and the evolution of resource constraints. The integrated framework not only significantly improves decision-making speed and accuracy in trajectory planning but also demonstrates superior generalization and robustness in highly dynamic environments.

The MAIT model is composed of an embedding layer, attention layers, Mamba layers, and a trajectory output layer. Within the attention and Mamba layers, there are also Multi-Layer Perceptron (MLP) sublayers and Root Mean Square Normalization (RMSNorm) sublayers. The MLP sublayer enhances the representation capacity through linear transformations, while the RMSNorm sublayer normalizes the data distribution and stabilizes the training dynamics \cite{2017transformer}. The structure of the MAIT model is illustrated in Fig. \ref{fig:encoder}. The working process of the MAIT model can be described as follows:

\begin{figure}[htbp]
\centering
\includegraphics[width=10cm]{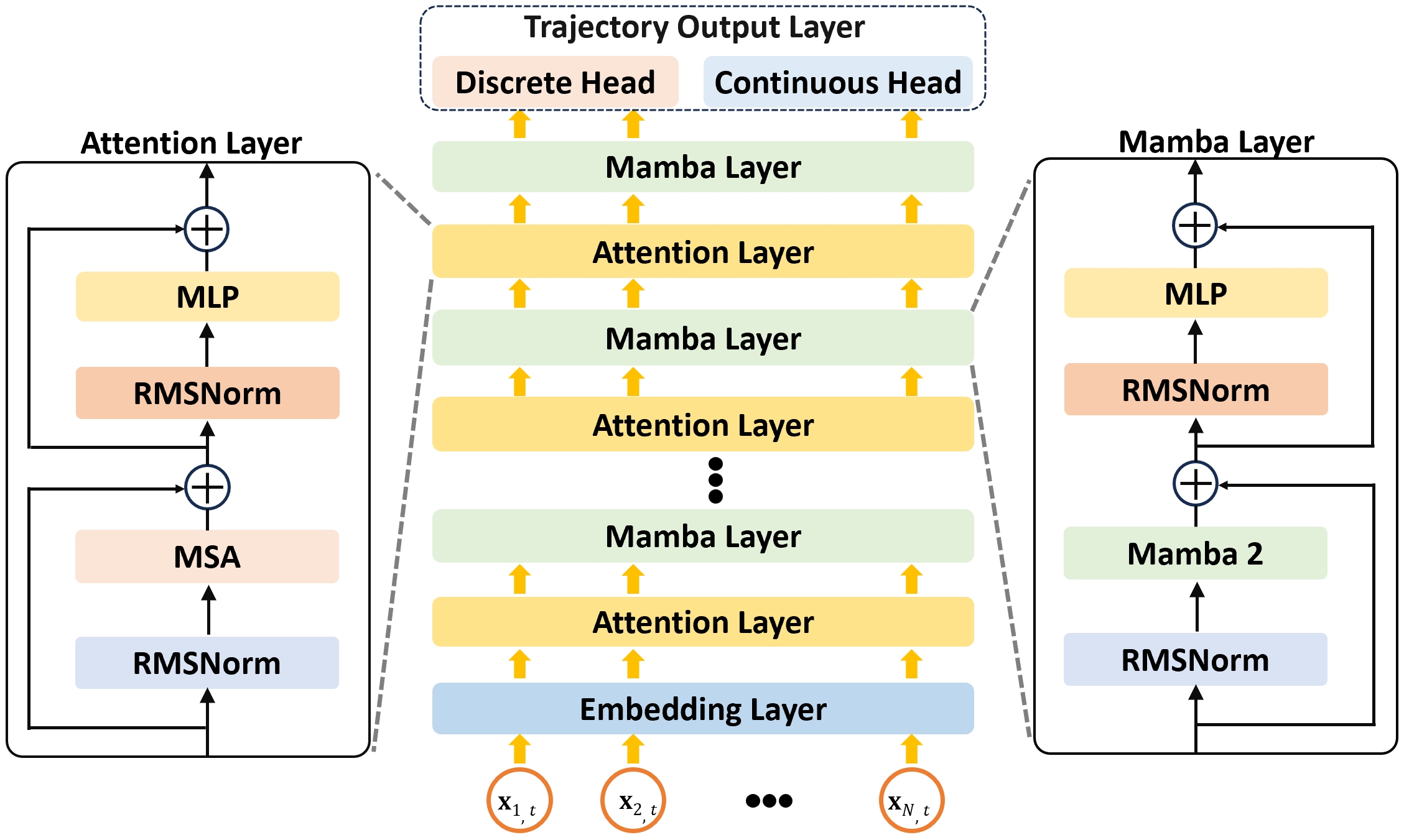}
\caption{Structure of the MAIT model.}
\label{fig:encoder}
\end{figure}

\subsubsection{Embedding Layer}
In the MAIT model, the efficient representation of input states is critical for capturing spatiotemporal dependencies and modeling collaborative behaviors among multiple UAVs. To address this, we design a structured embedding layer that maps the raw observation state of the UAV system at each time step into a unified high-dimensional representation, which serves as the input to the MAIT model. This module not only ensures dimensional alignment of the state information but also incorporates temporal awareness and individual identity modeling mechanisms, thereby enhancing the model’s capacity to represent sequential dynamics and heterogeneity.

Specifically, the input embedding layer consists of three components: (1) State embedding, (2) Temporal encoding, and (3) Identity embedding. The overall representation of the $i$-th input (either a ground device or a UAV) in the embedding layer is formulated as:
\begin{equation}
	\label{embedding}
	\textbf{h}_{i}^0 = \textbf{x}_{i, t} \textbf{W}_s + \textbf{p}_t + \textbf{b}_i
\end{equation}
where $\textbf{x}_{i, t} \in \mathbb{R}^{d_{\text{in}}}$ denotes the raw state vector of the $i$-th input at time step $t$, including features such as position, energy level, task load, and environmental observations.  
$\textbf{W}_s \in \mathbb{R}^{N \times d_{\text{in}}}$ is a linear projection matrix that maps the raw state into the dimensionality of the Transformer representations $d$.
$\textbf{p}_t \in \mathbb{R}^d$ is the temporal encoding vector for time step $t$, providing the necessary sequential information for sequence modeling, which is learnable.  
$\textbf{b}_i \in \mathbb{R}^d$ is the identity embedding of the $i$-th input, encoding its uniqueness in the task.

This embedding layer offers several advantages. First, the state embedding ensures unified alignment of input features from different attribute spaces, facilitating effective sequence modeling by the MAIT. Second, the temporal encoding equips the model with the capability to capture the sequential dynamics of trajectory evolution. Finally, the identity embedding explicitly models the heterogeneity and parallelism of ground devices and UAVs, providing a structural foundation for subsequent interaction-based attention mechanisms. Therefore, this embedding layer not only enhances the expressive power of the input representations but also provides robust structural priors to support multi-UAV trajectory planning tasks.

\subsubsection{Attention Layer}
The $l$-th attention layer of the MAIT maps the features of the $i$-th input $\textbf{h}_i^{l-1}$ to query $\textbf{Q}_i$, key $\textbf{K}_i$ and value $\textbf{V}_i$ with weights $\textbf{W}^q, \textbf{W}^k,\textbf{W}^v$, respectively. The equation is given as
\begin{equation}
	\label{eq:qkv}
	\textbf{Q}_i=\textbf{W}^q \textbf{h}_i^{l-1},\quad \textbf{K}_i=\textbf{W}^k \textbf{h}_i^{l-1},\quad \textbf{V}_i=\textbf{W}^v \textbf{h}_i^{l-1}
\end{equation}
where $\textbf{Q}_i, \textbf{K}_i, \textbf{V}_i$ present the query, key and value of the $i$-th input, respectively. $\textbf{W}^q, \textbf{W}^k,\textbf{W}^v$ are learnable matrices that map from $\textbf{h}_i^{l-1}$ to $\textbf{Q}_i, \textbf{K}_i$ and $\textbf{V}_i$. 
Using the attention mechanism, one input node may receive information from the other input node. Hence, the compatibility of the $i$-th input node and the $j$-th input node is denoted as $\boldsymbol{\delta}_{ij}$, which is given as follows:
\begin{equation}
	\label{eq:compatiblity}
	\boldsymbol{\delta}_{ij}=\frac{\textbf{Q}_i^\text{T} \textbf{K}_j}{\sqrt{d_k}}
\end{equation}
where $d_k$ is the dimension of the key $\textbf{K}_i$.  Here, the purpose of dividing by $\sqrt{d_k}$ is to scale the compatibility $\boldsymbol{\delta}_{ij}$ to avoid poor performance of self-attention\cite{2017transformer}. 
Then, the self-attention feature of the $i$-th input node is given as
\begin{equation}
	\label{eq:oi}
	\textbf{Z}_i=\sum_{j=1}^N \text{Softmax}(\boldsymbol{\delta}_{ij}) \textbf{V}_j
\end{equation}
where $\textbf{V}_j$ represents the value of the $j$-th input node, and the $\text{Softmax}$ function can normalize the output and preserve the data characteristics. 

The Multi-head Self-Attention (MSA) mechanism allows the attention layers to jointly attend to information from different representation subspaces at different input nodes\cite{2017transformer}. 
In MSA, we have multiple queries $[\textbf{Q}_{i,1},...,\textbf{Q}_{i,H}]$, keys $[\textbf{K}_{i,1},...,\textbf{K}_{i,H}]$ and values $[\textbf{V}_{i,1},...,\textbf{V}_{i,H}]$ for $H$ heads. Then, the $H$ heads self-attention feature $[\textbf{Z}_{i,1},...,\textbf{Z}_{i,H}]$ of the $i$-th input node is calculated by equations (\ref{eq:qkv})-(\ref{eq:oi}), respectively. Next, we calculate the MSA feature of the $i$-th input node at the $l$-th self-attention layer as
\begin{equation}
	\label{eq:msa}
	\textbf{h}^l_i=\textbf{W}^z [\textbf{Z}_{i,1}, \textbf{Z}_{i,2},...,\textbf{Z}_{i,H}]
\end{equation}
where the $W^z$ is the learnable weight to merge $H$ heads self-attention features. 

\subsubsection{Mamba Layer}
Mamba-2 is an improved SSM-based neural architecture designed to efficiently model long-range dependencies in sequential data \cite{dao2024transformers}. Unlike the attention mechanism, which relies on pairwise token interactions, Mamba-2 formulates sequence modeling as continuous-time state-space dynamics, enabling linear time complexity with respect to sequence length and hardware-efficient implementation. It is particularly well-suited for long-sequence processing, low-latency inference, and memory-constrained environments.

The fundamental principle of the Mamba-2 is to model the output $y(t)$
 as the solution of a linear dynamical system driven by the input signal $x(t)$ via a hidden state $h(t)$. 
The system is governed by the following equations:
\begin{equation}
	h'(t) = \textbf{A} h(t) + \textbf{B} x(t) \label{eq:ssm_continuous_h_prime_final}
\end{equation}
\begin{equation}
	y(t) = \textbf{C} h(t) \label{eq:ssm_continuous_y_prime_final}
\end{equation}
where $h'(t)$ is the derivative of $h(t)$ with respect to time $t$, $\textbf{A}$ is the learnable state matrix, $\textbf{B}$ is the learnable input matrix, and $\textbf{C}$ is the learnable output matrix.

Mamba-2 can be regarded as a discrete version of continuous SSMs. The discretization formulas are represented as follows:
\begin{equation}
	\mathbf{\bar{A}} = \exp(\Delta \textbf{A}) \label{eq:discrete_A_bar_prime_final}
\end{equation}
\begin{equation}
	\mathbf{\bar{B}} = (\Delta \textbf{A})^{-1}(\exp(\Delta \textbf{A})-\textbf{I})\cdot \Delta \textbf{B} \label{eq:discrete_B_bar_prime_final}
\end{equation}
where $\mathbf{\bar{A}}$ is the discretized learnable state matrix, $\mathbf{\bar{B}}$ is the discretized learnable input matrix, $\Delta$ is the discretization time step parameter, $\exp(\cdot)$ is the matrix exponential function, and $\textbf{I}$ is the identity matrix.

Building upon this foundation, the continuous system equations are transformed into their discrete counterparts as follows:
\begin{equation}
		\label{eq:9}
	h_{t}=\mathbf{\bar{A}}h_{t-1}+\mathbf{\bar{B}}x_{t} 
\end{equation}
\begin{equation}
	\label{eq:10}
	y_{t}=\textbf{C}h_{t} 
\end{equation}
where $h_t$, $h_{t-1}$, $x_t$, and $y_t$ are the current hidden state, previous hidden state, current input, and current output at discrete time step $t$, respectively;

In addition, for an input sequence with size $T$, a global convolution with kernel $\overline{\mathbf{K}}$ can be applied for computing the output of  Eqs. (\ref{eq:9})-(\ref{eq:10}) as in the following:
\begin{equation}
		\label{eq:12}
	\overline{\textbf{K}} = \left( {\textbf{C}}\overline{\textbf{B}}, \; {\textbf{C}}\overline{\textbf{A}}\overline{\textbf{B}}, \; \ldots, \; {\textbf{C}}\overline{\textbf{A}}^{T-1}\overline{\textbf{B}} \right)
\end{equation}
\begin{equation}
	\label{eq:13}
	\mathbf{y} = \textbf{x} * \overline{\textbf{K}},
\end{equation}
where $\textbf{x}$ and $\mathbf{y}$ denote the input and output matrices of the Mamba layer, respectively.

The Mamba-2 layers are inherently designed to capture diverse contextual information through selective state propagation, eliminating the need for the explicit multi-head architecture typically employed in Transformer models.

\subsubsection{Trajectory Output Layer}
In UAV trajectory optimization tasks, the trajectory output layer serves as the terminal module of the Transformer architecture, responsible for decoding the high-dimensional temporal representations generated by the MAIT model into executable actions or trajectory control signals. The design of the output layer directly determines the mapping between the model and real-world control tasks, functioning as a critical bridge between the model’s representational space and the actual flight control space. Therefore, developing a structurally clear, functionally complete, and task-generalizable output mechanism, capable of handling both discrete and continuous trajectory planning modes, is essential for enhancing the system’s adaptability and generalization across diverse UAV scenarios.

The trajectory output layer is designed to support the simultaneous modeling and parallel output for multiple UAVs, aiming to handle both discrete trajectory planning and continuous trajectory control tasks. Its fundamental function is to decode the UAV trajectories from the complete feature sequence representations generated by the Transformer with $L$ layers, denoted as
$\textbf{H}_t = \{ \textbf{h}_{1,t}^{L}, \textbf{h}_{2,t}^{L}, \ldots,\textbf{h}_{N,t}^{L} \} \in \mathbb{R}^{N \times d}$
into the corresponding action outputs
${\textbf{A}}_t = \{ \textbf{{a}}_{1,t}, \textbf{{a}}_{2,t}, \ldots, \textbf{{a}}_{M,t} \} \in \mathbb{R}^{M \times K}$, where $N$ is the number of input nodes, $d$ is the dimensionality of the Transformer representations, $M$ is the number of UAVs, and $K$ is the dimensionality of the task-related output space.

The discrete trajectory head is designed to handle trajectory planning tasks where the set of target nodes consists of a finite number of stop points. For example, in scenarios involving data collection from fixed IoT devices, the trajectory head enables each UAV to select the most appropriate target from a predefined set of stop points $\mathcal{V} = \{ v_1, \ldots, v_K \}$. The output is a probability distribution over the candidate stop points for each UAV at the $t$-th time step:
\begin{equation}
	\label{eq:prob}
	p_j(t)=\text{Softmax}(\textbf{W}^{Q} \mathbf{H}_t) \textbf{M}
\end{equation}
where  $\textbf{W}^{Q}$ is the learnable projection matrix,
and $\textbf{M}$ is the mask matrix. Initially, we set $\textbf{M}=[1,...,1]$ at the beginning of trajectory planning. The length of $\textbf{M}$ is $K$. When $\textbf{M}(i)=1$ means UAV can fly to the $i$-th stop point,, while $\textbf{M}(i) = 0$ means that the stop point is currently inaccessible or restricted. 

In each time step, each UAV selects the stop point with max probability in $\textbf{P}(t)=\left\{p_j(t),j\in \mathcal{V} \right\}$ as the next stop point, so one has
 \begin{equation}
\label{eq:next}
\textbf{A}_t= \text{argmax} (\textbf{P}(t))
\end{equation}
where $\text{argmax}(\cdot)$ returns the index of the stop point with the max probability.

The continuous trajectory head is applied to scenarios that require the output of continuous action variables, such as position coordinates, orientation vectors, and similar representations. Typical applications include smooth trajectory following or dynamic navigation. Specifically, the predicted action $\textbf{{a}}_{i,t}$ can take various formats depending on the control task, such as:
$(\Delta x, \Delta y)$ representing position increments, $(d, \theta)$ representing distance and heading angle,
or $(x, y)$ representing the absolute target position, so one has
\begin{equation}
	\label{eq:prob2}
\textbf{A}_t=\text{Tanh}(\textbf{W}^{Q} \mathbf{H}_t) \textbf{S}
\end{equation}
where  $\textbf{W}^{Q}$ is the learnable projection matrix,
and $\textbf{S}$ is the scale matrix, which is a task-specific physical scaling factor used to adjust the predicted output to a physically meaningful range.

The proposed output layer exhibits several critical functionalities that enhance its suitability for multi-UAV trajectory planning tasks. First, it supports parallel modeling across multiple UAVs by enabling unified state representation and simultaneous output generation, thereby capturing coordination more effectively. Second, the architecture demonstrates strong adaptability to diverse task formats by incorporating distinct output heads for discrete selection and continuous control tasks, facilitating seamless integration across heterogeneous mission requirements. Third, by leveraging Softmax and Tanh activation functions for discrete and continuous action spaces, respectively, the framework achieves a balance between output precision and control stability, contributing to more robust policy learning in RL-based optimization. The workflow of the MAIT model is presented in \textbf{Algorithm \ref{alg:mait}}.

\begin{algorithm}[htbp]
	\caption{Mamba-Attention Integrated Transformer}
	\label{alg:mait}
	\begin{algorithmic}[1]
		\REQUIRE Raw state observations at each time step.
		\ENSURE Trajectory outputs for all UAVs.
		\STATE Initialize the MAIT model parameters		
		\FOR{each time step $t$}
		\FOR{each input $i$}
		\STATE Map raw state to embedding features by Eq. (\ref{embedding}).
		\ENDFOR
		
		\FOR{each encoder layer $l$}
		\IF{layer type is attention layer}
		\STATE Compute MSA by Eqs. (\ref{eq:qkv})-(\ref{eq:msa}).
		\ELSE
		\STATE Compute Mamba-2 state-space transformation by Eqs. (\ref{eq:12})-(\ref{eq:13}).
		\ENDIF
		\STATE Apply the RMS normalization and MLP transformation.
		\ENDFOR
		
		\IF{discrete trajectory planning}
		\STATE Use Softmax function to predict probability over candidate stop points and select next target by Eqs. (\ref{eq:prob})-(\ref{eq:next}). 
		\ELSE
		\STATE Use Tanh function to predict continuous control variables by Eq. (\ref{eq:prob2}).
		\ENDIF
		
		\ENDFOR
		
		\RETURN Final planned trajectories for all UAVs.
	\end{algorithmic}
\end{algorithm}

\subsection{Trajectory-Group Relative Policy Optimization}
\label{sec:train}
Conventional RL methods for UAV trajectory planning primarily rely on Actor-Critic (AC)-type algorithms, such as A3C and DDPG. However, these methods face significant challenges during training. On one hand, AC algorithms require the simultaneous maintenance of multiple policy networks and value networks. The value network typically has a scale comparable to the policy network, resulting in considerable computational and memory overhead. On the other hand, in highly dynamic environments, the value of intermediate states is difficult to estimate accurately, leading to instability during training.

To address the challenges of resource consumption and optimization instability associated with AC-based training in UAV trajectory planning, we propose a novel T-GRPO method. This method completely eliminates the need for a value network, significantly reducing computational and memory costs during RL. Moreover, T-GRPO estimates advantages based on the relative quality among multiple generated outcomes under the same state, rather than relying on predicted rewards from a value network, thereby mitigating estimation bias. In addition, T-GRPO introduces an optimized regularization strategy for Kullback–Leibler (KL) divergence, further enhancing training stability and convergence efficiency. 

\begin{figure}[htbp]
	\centering
	\includegraphics[width=7cm]{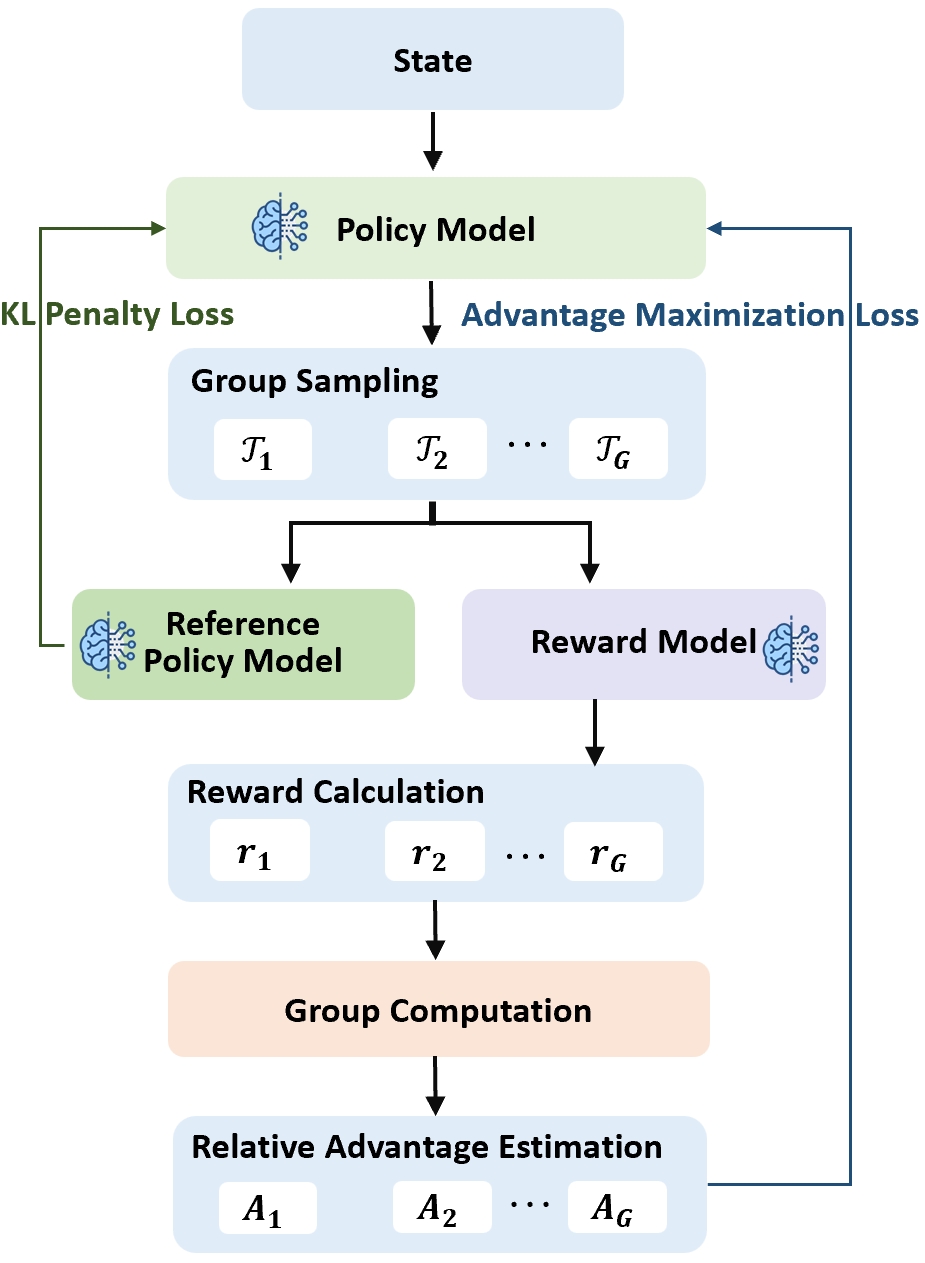}
	\caption{Training procedure of T-GRPO.}
	\label{fig:GRPO}
\end{figure}

In the following, we present a detailed formulation of the T-GRPO training procedure, as illustrated in Fig. \ref{fig:GRPO}, beginning with the definitions of the state, action, and reward.

\subsubsection{State, Action and Reward} 

To systematically formalize the multi-UAV cooperative trajectory planning task, the definitions are given as follows:

\begin{itemize}
	\item \textbf{State space $s_t$:}
	The system state at time step $t$ consists of the joint states of all input nodes, expressed as
\begin{equation}
	s_t = \left\{ s_{1,t}, s_{2,t}, \dots, s_{N,t} \right\}, \quad s_{i,t} \in \mathbb{R}^{d_{\text{in}}}
\end{equation}
	where $s_{i,t}$ denotes the state vector of the $i$-th input node at time $t$.
	
	\item \textbf{Action space $a_t $:}
	Depending on the specific task formulation, the action space is defined as follows:
\begin{equation}
	a_t = \left\{ a_{1,t}, a_{2,t}, \dots, a_{M,t} \right\}, \quad a_{i,t} \in \mathbb{R}^{K}
\end{equation}	
	
For discrete planning tasks, each UAV selects its next target from a predefined set of candidate stop points denoted by 
$\mathcal{V} = \{ v_1, \ldots, v_K \}$. In the case of continuous control, each UAV instead generates a continuous control vector to adjust its motion parameters. Hence, the joint RL trajectory of the system over a planning horizon is represented by the sequence of states and actions:
\begin{equation}
\tau = \left\{ s_t, a_t \right\}_{t=0}^{T}
\end{equation}
where $T$ is the time horizon of the planning process.

\item \textbf{Reward function $r(s_t, a_t)$:}
The reward function reflects the instantaneous system benefit, such as negative energy consumption, coverage gain, path smoothness, or collision avoidance penalties. The cumulative discounted return over the entire RL trajectory is defined as
\begin{equation}
R(\tau) = \sum_{t=0}^{T} \gamma^{t} \, r(s_t, a_t)
\end{equation}
where $\gamma \in [0,1]$ is the discount factor, $r(\cdot)$ is the task-defined reward function.
\end{itemize}

\subsubsection{Group Sampling}
To sufficiently explore the policy space and mitigate the bias introduced by a single trajectory, a multi-trajectory sampling strategy is employed. In each round of policy updates, the algorithm starts from a given initial state $s_0$ and performs $G$ independent rollouts under the current policy $\pi_\theta$, thereby generating $G$ complete sequences of state-action pairs. Each rollout fully records the transitions and action selections from the initial state to the terminal step, which can be formalized as
\begin{equation}
\tau^{(j)} = \left\{ s_0, a^{(j)}_{0}, s^{(j)}_{1}, a^{(j)}_{1}, \dots, s^{(j)}_{T}, a^{(j)}_{T} \right\}, \quad j=1,2,\dots,G.
\end{equation}

This multi-trajectory parallel sampling scheme enhances the diversity of policy behavior representation and ensures the robustness and sufficiency of subsequent advantage estimation.

\subsubsection{Reward Calculation}
The cumulative return of each trajectory is obtained by summing the discounted instantaneous rewards over time. Specifically, with a discount factor $\gamma \in [0,1]$, the cumulative reward of the $j$-th trajectory is defined as
\begin{equation}
	\label{eq:reward}
R^{(j)} = \sum_{t=0}^{T} \gamma^{t} \, r\left( s_{t}^{(j)}, a_{t}^{(j)} \right).
\end{equation}

This cumulative reward captures the overall performance of UAVs during the planning horizon.

\subsubsection{Relative Advantage Estimation}
To avoid bias from absolute value estimation, T-GRPO adopts a normalization-based relative advantage across the group. The mean $\mu_R$ and standard deviation $\sigma_R$ of the group returns are calculated as
\begin{equation}
	\label{eq:mean}
\mu_R = \frac{1}{G} \sum_{j=1}^{G} R^{(j)}
\end{equation}
\begin{equation}
	\label{eq:std}
\quad
\sigma_R = \sqrt{ \frac{1}{G} \sum_{j=1}^{G} \left( R^{(j)} - \mu_R \right)^2 + \delta }
\end{equation}
where $\delta$ is a small positive constant for numerical stability. The relative advantage of the $j$-th trajectory is then expressed as
\begin{equation}
		\label{eq:advantage}
A^{(j)} = \frac{ R^{(j)} - \mu_R }{ \sigma_R }.
\end{equation}

This mechanism improves the stability of policy optimization and mitigates the impact of varying reward scales during updates.

\subsubsection{Policy Optimization}

In the T-GRPO training process, for each trajectory and each decision step $t$, the probability ratio between the current policy and the old policy is first computed to measure the relative change in action distributions, formulated as
\begin{equation}
		\label{eq:26}
r_{t}^{(j)} = \frac{ \pi_\theta\left( a_{i,t}^{(j)} \mid s_t \right) }{ \pi_{\theta_\text{old}}\left( a_{i,t}^{(j)} \mid s_t \right) }
\end{equation}
where $\pi_\theta$ and $\pi_{\theta_\text{old}}$ denote the probability distributions of the current and previous policies, respectively.

To guarantee stable policy updates, the T-GRPO framework adopts a clipped objective inspired by PPO \cite{schulman2017proximal}, constraining the excessive amplification of relative advantage. The GRPO loss at step $t$ on the $j$-th trajectory is defined as
\begin{equation}
L_t^{(j)} = \min \left( r_t^{(j)} A^{(j)}, \, \text{clip}\left(r_t^{(j)}, 1 - \epsilon, 1 + \epsilon\right) A^{(j)} \right)
\end{equation}
where $\text{clip}(\cdot)$ is a clipping function, which serves to constrain the policy probability ratio $r_t^{(j)}$ in a controllable range, thereby preventing overly aggressive updates that may cause instability, and $\epsilon$ is a hyperparameter controlling the clipping range.

The overall policy objective is then expressed as maximizing the average across all sampled trajectories and time steps:
\begin{equation}
	\label{eq:28}
\mathcal{L}(\theta) = - \frac{1}{G} \sum_{j=1}^{G} \sum_{t=0}^{T} L_t^{(j)}.
\end{equation}

Furthermore, to regulate the magnitude of policy shifts and prevent excessive divergence during updates, a KL penalty term is introduced, yielding the total objective function as follows:
\begin{equation}
		\label{eq:29}
\mathcal{L}_\text{total} = \mathcal{L}(\theta) + \lambda \frac{1}{G} \sum_{j=1}^{G} \sum_{t=0}^{T} D_\text{KL} \left( \pi_{\theta_\text{old}}(\cdot \mid s_t) \,\|\, \pi_\theta(\cdot \mid s_t) \right)
\end{equation}
where $\lambda$ is a regularization coefficient. This composite objective effectively balances policy improvement, update stability, and group-relative advantage, thereby enhancing convergence speed and robustness in multi-UAV trajectory planning tasks. The workflow of the T-GRPO algorithm is presented in \textbf{Algorithm \ref{alg:tgrpo}}.

\begin{algorithm}[htbp]
	\caption{Trajectory-Group Relative Policy Optimization}
	\label{alg:tgrpo}
	\begin{algorithmic}[1]
		\REQUIRE Initial policy $\pi_\theta$, initial state $s_0$, total rollout episodes $G$, time horizon $T$.
		\ENSURE Optimized policy parameters $\theta$.
		
		\WHILE{not converged}
		\STATE Sample $G$ independent trajectories starting from $s_0$ under $\pi_\theta$.
		
		\FOR{each trajectory $j=1,\dots,G$}
		\STATE Compute cumulative reward by Eq. (\ref{eq:reward}).
		\ENDFOR
		
		\STATE Compute group mean and standard deviation by Eqs. (\ref{eq:mean})-(\ref{eq:std}).
		\FOR{each trajectory $j=1,\dots,G$}
		\STATE Calculate relative advantage by Eq. (\ref{eq:advantage}).
		\ENDFOR

		\STATE Compute the overall policy objective by Eqs. (\ref{eq:26})-(\ref{eq:28}).
		
		\STATE Compute the total loss $\mathcal{L}_\text{total}$ by incorporating KL divergence regularization in Eq. (\ref{eq:29}).
		
		\STATE Update policy $\pi_\theta$ by minimizing $\mathcal{L}_\text{total}$.
		
		\ENDWHILE
		
		\RETURN Optimized policy parameters $\theta$.
	\end{algorithmic}
\end{algorithm}

\section{Case Study 1. Continuous Trajectory Planning}
\label{sec:experiments1}
\subsection{Simulation Parameter Setting}

In this case study, we investigate the trajectory planning problem in a UAV-assisted Mobile Edge Computing (MEC) system, adopting a system framework similar to that proposed in \cite{wang2021deep}. The simulation area covers 400 × 400 meters, with 200 User Equipments (UEs) and four UAVs. Each UE generates one task in each time slot, where the communication data size of each task is selected between 10 and 50 KB, and the computational resource demand is chosen within the range of $2 \times 10^{9}$ to $2 \times 10^{10}$ CPU cycles. The remaining parameters of this case study are detailed in Table \ref{tab:param1} \cite{wang2021deep}.

\begin{table}[h]
	\caption{Parameter settings for case study 1}
	\centering
	\setlength{\tabcolsep}{4mm}
	\renewcommand{\arraystretch}{1.3}
	\begin{tabular}{|p{5cm}|p{2cm}|}\hline
		Parameter&{Value}\\\hline
		Number of time slots $T$ &60\\
		
		Number of UEs $N$&100\\
		Number of UAVs $M$& 4\\
		Maximal number of tasks $V^{max}$&30\\
		Maximal distance $d^{max}$& 30 m\\
		Maximal time duration $T^{max}$& 1 s\\
		Maximal azimuth angle $\theta^{max}$& ${\pi}/{4}$\\
		Altitude value of the UAV $Z_{j}(0)$&75 m\\
		Channel power gain $g_{0}$& $1.42 \times 10^{-4}$\\
		Transmitting power $P^{Tr}$& 0.1 W\\
		Channel bandwidth $B$& 10 MHz\\
		Noise power $\sigma^{2}$&-90 dbm\\
		Maximal energy of each UAV $e^{max}$& $10^{6}$ J\\
		Effective switched capacitance $k_i$& $10^{-28}$\\
		Maximal computation resource $f^{max}$& 100 GHz\\
		\hline
	\end{tabular}
	
	\label{tab:param1}
\end{table}

\subsection{Agentic RAG Analysis}
We designed an Agentic RAG system based on DeepSeek R1, harnessing its powerful reasoning capabilities to better support agent-driven optimization of the RAG framework. We adopted Qanything as the underlying RAG system \cite{jiang2025commgpt} and enriched its knowledge base with a substantial amount of literature related to UAV trajectory planning, including in particular reference \cite{wang2021deep}, to enhance the LLM’s understanding of UAV mission requirements. After multiple rounds of retrieval, the LLM provided the following conclusions: 

\subsubsection{State space} The state space consists of the positions of the UEs, along with the communication data size and computational resource demand of each task.

\subsubsection{Action space} The action space is defined as the flight distance and angle of each UAV. It is assumed here that the task offloading matrix and the resource allocation matrix can be addressed using existing convex optimization or matching algorithms \cite{wang2021deep}, which are available in the tool library of the Agentic RAG framework.

\subsubsection{Reward function} The reward function is designed to minimize the total energy consumption of all UEs, including both task offloading energy and local execution energy.

\subsubsection{Model architecture} The recommended model architecture consists of 10 encoder layers, comprising 70\% attention layers and 30\% Mamba layers.

\subsubsection{Training method}
The training method employs a continuous trajectory optimization strategy with 4 independent rollouts to train the model.

\subsubsection{Task constraints} The task constraints include binary offloading decisions for each UE’s tasks, the maximum number of offloaded tasks that can be executed by a single UAV, UAV flight range limitations, UAV coverage constraints, and quality-of-service requirements for task execution.

\subsection{Comparison of Different Transformer Models}

This experiment is presented to compare the performance and efficiency of different network architectures.
The MAIT model is compared with the following attention mechanisms: Classic Transformer \cite{vaswani2017attention}, Linear Transformer (Linformer)\cite{wang2020linformer}, and Performer \cite{choromanski2020rethinking}. We tested two different MAIT models: the MAIT model (MAIT 1) recommended by the Agentic RAG system incorporated 70\% attention layers, while another MAIT model (MAIT 2) incorporated 30\% attention layers.
The energy consumption and inference time of different Transformer models are shown in Table \ref{tab:perform}. 

\begin{table}[h]
	\caption{Comparison of different Transformer models}
	\centering
	\setlength{\tabcolsep}{3mm}
	\renewcommand\arraystretch{1.3}
	  \begin{tabular}{|l|c|c|}
		\hline
		Method & Energy Consumption (J) & Inference Time (s) \\
		\hline
			Transformer  \cite{vaswani2017attention} & 376.38 & 2.25 \\
	Linformer\cite{2015luong} & 398.65 & 1.54 \\
		Performer\cite{2019area} & 338.43 & 1.66 \\
		MAIT  1 & \textbf{317.64} & 1.42 \\
		MAIT 2 & 340.57 & \textbf{1.15} \\
		\hline
	\end{tabular}

	\label{tab:perform}
\end{table}

As shown in Table \ref{tab:perform}, it is evident that the MAIT 1 model recommended by the Agentic RAG system achieves the lowest energy consumption. The MAIT 2 model, which incorporates more Mamba layers, achieves the shortest inference time but at the cost of slightly increased energy consumption. The Performer model demonstrates lower energy consumption than MAIT 2, but with a longer inference time. Other Transformer-based models do not exhibit advantages in either performance or inference time compared to the MAIT models.

The reasons for the low energy consumption can be interpreted as follows:
(1) The Agentic RAG framework can determine an appropriate MAIT architecture (i.e., the proportion of attention and Mamba layers) depending on the user scale, thereby achieving efficient and high-accuracy feature extraction.
(2) The hybrid structure of attention and Mamba enables the model to capture both long-range dependencies and fine-grained local patterns simultaneously. Specifically, the attention layers excel at modeling global contextual relationships, while the Mamba layers effectively preserve sequential consistency and reduce redundant computations.

\subsection{Comparison of Different Training Methods}
This experiment aims to investigate the feasibility of the proposed T-GRPO algorithm. The T-GRPO algorithm is compared with the following RL methods:  AC \cite{konda1999actor}, DDPG \cite{tan2021reinforcement}, and SAC \cite{haarnoja2018soft}. 
The energy consumption and training time of different RL methods are listed in Table \ref{tab:baseline}. As shown in Table \ref{tab:baseline}, T-GRPO achieves the lowest energy cost among all algorithms. This can be attributed to the T-GRPO algorithm, which incorporates a normalized advantage function design to significantly reduce the variance in policy gradients, thereby enabling more stable and efficient optimization of the energy consumption objective. In addition, T-GRPO dynamically adjusts the balance between exploration and exploitation, allowing it to converge to a more optimal energy control policy in diverse state spaces. As a result, it demonstrates more robust and superior performance during the training phase.


\begin{table}[h]
	\caption{Comparison of different training methods}
	\centering
	\setlength{\tabcolsep}{2mm}
	\renewcommand\arraystretch{1.3}
	\begin{tabular}{|l|c|c|}\hline
		Energy Consumption&Energy Consumption (J) & Training Time (s)\\\hline
		AC\cite{konda1999actor}&493.21&13104.46\\
		DDPG \cite{tan2021reinforcement}&457.79&10778.73\\
		SAC\cite{haarnoja2018soft}& 349.04& \textbf{9905.53}\\
		T-GRPO& \textbf{314.53}& 10404.27\\
		\hline
	\end{tabular}
	\label{tab:baseline}
\end{table}

\section{Case Study 2. Discrete Trajectory Planning}
\label{sec:experiments2}

\subsection{Simulation Parameter Setting}
In this case study, we examine the trajectory planning of UAVs in a Wireless Power Transfer (WPT)-assisted Internet of Things (IoT) system, adopting a framework similar to that presented in \cite{10879146}. The study area encompasses a region measuring 1000 × 1000 meters, in which 500 IoT Devices (IoTDs) are randomly deployed. The amount of data collected by each IoTD is randomly drawn from the interval between 0.2 and 1.5 MB. Then, the detailed parameter settings of the case study are listed in Table \ref{tab:param2}\cite{10879146}. 

\begin{table}[h]
	\caption{Parameter settings for case study 2}
\centering
\setlength{\tabcolsep}{8mm}
\renewcommand{\arraystretch}{1.3}
	\begin{tabular}{|l|l|}\hline
		Parameter&{Value}\\\hline
		Number of IoTDs $N$&500\\
		Number of UAVs $M$& 4\\
		UAV Data Storage Capacity $C_{\max}$&150 MB\\
		Data Transfer Rate $R_{ij}$&1024 KB/s\\
		Data Transmitting Power $P^T$&50 W\\
		Channel power gain at 1m $\kappa_{0}$ & -60 dB\\
		Path loss factor $\kappa$ & 2\\
		UAV Battery Capacity $E_{\max}$&2550 mAh\\      
		Harvested Power $\phi(P^{R}_{ij})$ &50 W\\
		Fly speed of UAV $v$ & 10 m/s\\
		Flight Height $H^F$ & 20 m\\
		Flight Power $P^F$&75 W\\
		Bandwidth $B$ & 2M Hz\\
		Transmitting power $P^{T}$ & 0.5 W\\
		Data collection power $P^{C}$ & 0.5 W\\
		Hover power $P^{H}$ & 50 W\\
		Noise power $\sigma^{2}$ & -110 dBm\\
		\hline
	\end{tabular}
	
	\label{tab:param2}
\end{table}


\subsection{Agentic RAG Analysis}

We incorporated the reference work \cite{10879146} into the knowledge base and constructed queries for the LLM concerning a similar system. After multiple rounds of retrieval, the LLM provided the following conclusions: 

\subsubsection{State space} The state space consists of the locations of all IoTDs along with their remaining data to be collected.

\subsubsection{Action space}  The action space is defined as selecting the location of the next IoTD to be visited, under the constraint that each IoTD can be visited only once and exclusively by a single UAV.

\subsubsection{Reward function} The reward function accounts for the total UAV energy consumption, including flight energy, hovering energy, and wireless charging energy.

\subsubsection{Model architecture} The recommended model architecture consists of 15 encoder layers, comprising 30\% attention layers and 70\% Mamba layers, with a suggested discrete output head incorporating a masking matrix.

\subsubsection{Training method}
The training method employs a discrete trajectory optimization strategy with 8 independent rollouts to train the model.

\subsubsection{Task constraints}  The task constraints stipulate that, due to the limited data storage and energy resources of UAVs, the total energy consumption must not exceed the UAV’s maximum energy capacity, and the total amount of collected data must remain within its maximum storage capacity.

\subsection{Comparison of Different Trajectory Designers}
This experiment aims to compare the overall performance of the proposed MAIT model with different discrete trajectory planning algorithms. Specifically, the MAIT model is compared with Graph Neural Network (GNN) \cite{2017-cvrp-gnn}, Graph Pointer Network (GPN) \cite{2019-gpn-rl}, and ATOM network \cite{10879146} in terms of energy consumption and inference time, as summarized in Table \ref{tab:method}.

As shown in Table \ref{tab:method}, the MAIT model achieves the lowest energy consumption and the shortest inference time among all compared algorithms. The reasons can be explained as follows: the MAIT framework leverages a hybrid Attention and Mamba mechanism, which allows the encoder to capture both local and global IoTD features for more effective feature representation; meanwhile, the Mamba structure can significantly reduce the inference time.

\begin{table}[h]
	\caption{Comparison of different trajectory designers}
	\centering
	\setlength{\tabcolsep}{4mm}
	\renewcommand\arraystretch{1.4}
	\begin{tabular}{|l|c|c|}\hline
		Methods&{Energy Consumption (Wh)}&{Inference Time (s)}\\\hline
		GNN\cite{2017-cvrp-gnn}&292.861&2.42\\
		GPN\cite{2019-gpn-rl}&278.88&1.67\\
		ATOM \cite{10879146}&260.64&2.86\\
		MAIT &\textbf{249.57}& \textbf{1.53}\\
		\hline
	\end{tabular}
	\label{tab:method}
\end{table}

\subsection{Comparison of Different Constraints}

\begin{figure}[htbp]
	\centering
	\includegraphics[width=9cm]{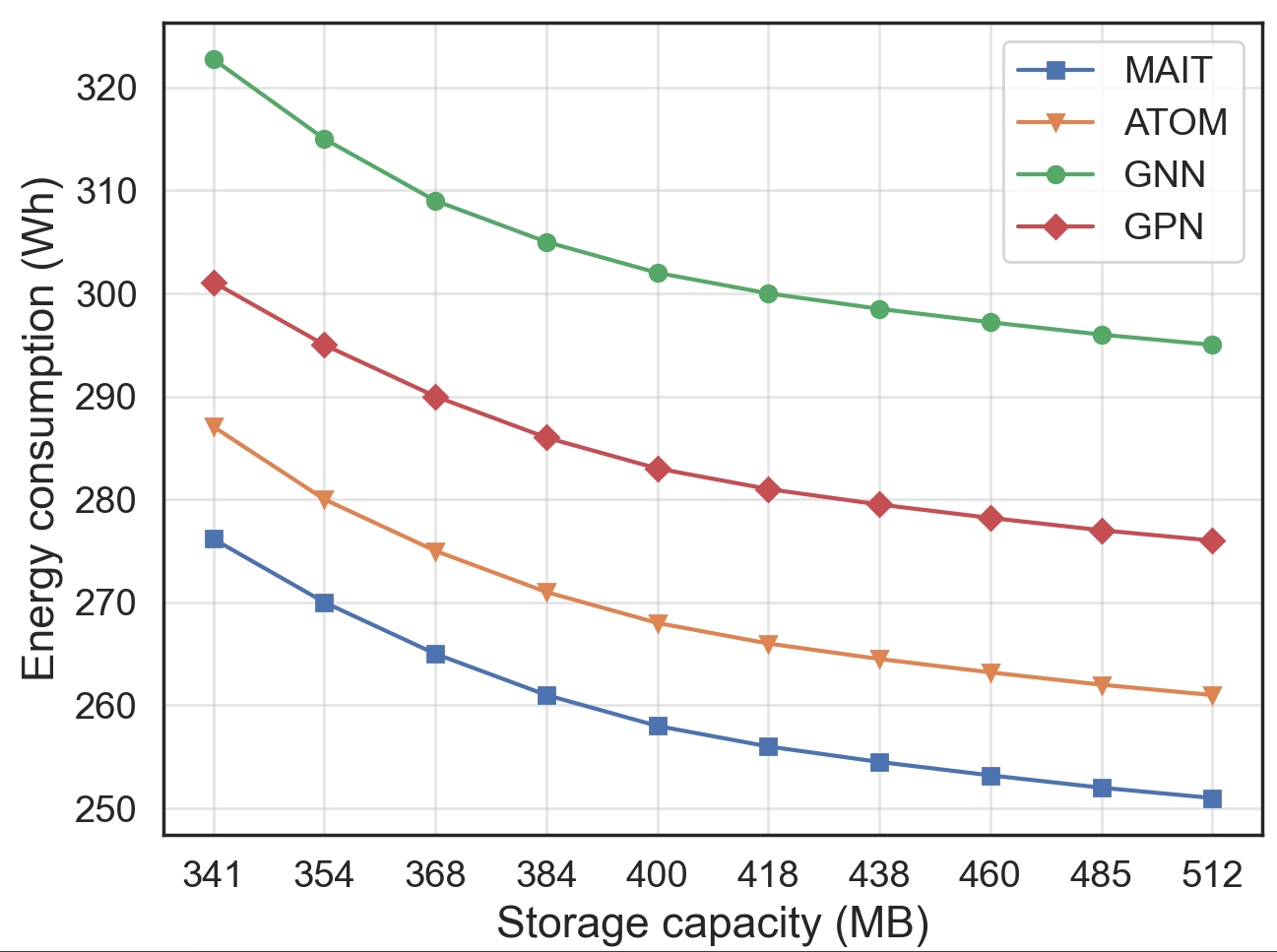}
	\caption{Energy cost with different data storage capacities of UAVs.}
	\label{fig:change_data}
\end{figure}

This experiment aims to evaluate the generalization of the MAIT model, and we apply all trajectory designers to scenarios with different UAV data storage capacities. The range of UAV data storage capacity is set from $340$ to $512$ MB. The simulation results of the energy costs are shown in Fig. \ref{fig:change_data}.

As shown in Fig. \ref{fig:change_data}, it can be observed that GNN incurs the highest energy cost, followed by GPN, ATOM, and MAIT.
It is evident that the MAIT model achieves the lowest energy cost among all neural network-based methods. The reasons why the MAIT model outperforms other neural network-based approaches can be summarized as follows:
(1) The hybrid self-attention and Mamba mechanisms in the MAIT model exhibit more powerful feature extraction capabilities than other neural network-based methods;
(2) The proposed T-GRPO can better learn and adapt to the dynamic elements in UAV-assisted data collection systems, such as remaining data storage and battery capacity. Therefore, the MAIT model can fully leverage the constraints of data storage to generate the optimal UAV trajectory with the lowest energy cost.

\section{Conclusions}
\label{sec:conclusion}
This paper proposed the ARMAIT framework, which integrates Agentic RAG for autonomous task understanding, the MAIT model for efficient trajectory feature modeling, and T-GRPO for stable RL optimization across mixed action spaces. ARMAIT enables end-to-end reasoning and trajectory generation for multi-UAV systems. Experimental results show that ARMAIT achieves superior decision-making accuracy and faster convergence than existing approaches and demonstrates strong generalization across multiple case studies.

However, the current framework still relies on global system information and only optimizes UAV trajectories, without joint consideration of resource scheduling. In the future, we will extend ARMAIT to decentralized and partially observable settings and explore joint optimization of trajectory and resource allocation to enhance overall system performance.

\bibliographystyle{IEEEtran}
\bibliography{bare_jrnl}
\newpage
\end{document}